\documentclass{ws-procs975x65}

\begin{document}

\title{Magnetic monopole versus vortex as gauge-invariant topological objects for quark confinement}

\author{Kei-Ichi Kondo$^\#$\footnote{Speaker at the workshop SCGT15.}, Takaaki Sasago, Toru Shinohara}

\address{Department of Physics,  
Graduate School of Science, 
Chiba University, Chiba 263-8522, Japan\\
%$^*$E-mail: kondok@faculty.chiba-u.jp
}

\author{Akihiro Shibata}

\address{
%Computing Research Center,  
High Energy Accelerator Research Organization (KEK),  
%and Graduate Univ. for Advanced Studies (Sokendai), 
Tsukuba  305-0801, Japan\\
%E-mail: akihiro.shibata@kek.jp
}

\author{Seikou Kato}
\address{Fukui National College of Technology, Sabae 916-8507, Japan\\
%E-mail: skato@fukui-nct.ac.jp
}

%\author{Toru Shinohara}
%\address{Department of Physics,  
%Graduate School of Science, 
%Chiba University, Chiba 263-8522, Japan\\
%E-mail: sinohara@graduate.chiba-u.jp}

\begin{abstract}
First, we give a gauge-independent definition of chromomagnetic monopoles in $SU(N)$ Yang-Mills theory which is derived through a non-Abelian Stokes theorem for the Wilson loop operator. Then we discuss how such magnetic monopoles can give a nontrivial contribution to the Wilson loop operator for understanding the area law of the Wilson loop average.
Next, we discuss how the magnetic monopole condensation picture are compatible with the vortex condensation picture as another promising scenario for quark confinement. We analyze the profile function of the magnetic flux tube as the non-Abelian vortex solution of $U(N)$ gauge-Higgs model, which is to be compared with numerical simulations of the $SU(N)$ Yang-Mills theory on a lattice.
This analysis gives an estimate of the string tension based on the vortex condensation picture, and possible interactions between two non-Abelian vortices. 
%Moreover, if time permits, we discuss a quantum origin of the scale generation due to dimensional transmutation using the novel large $N$ analysis for $SU(N)$ Yang-Mills theory.
\end{abstract}


\bodymatter

\section{Introduction}

\begin{figure}[ptb]
\begin{center}
\includegraphics[scale=0.45]{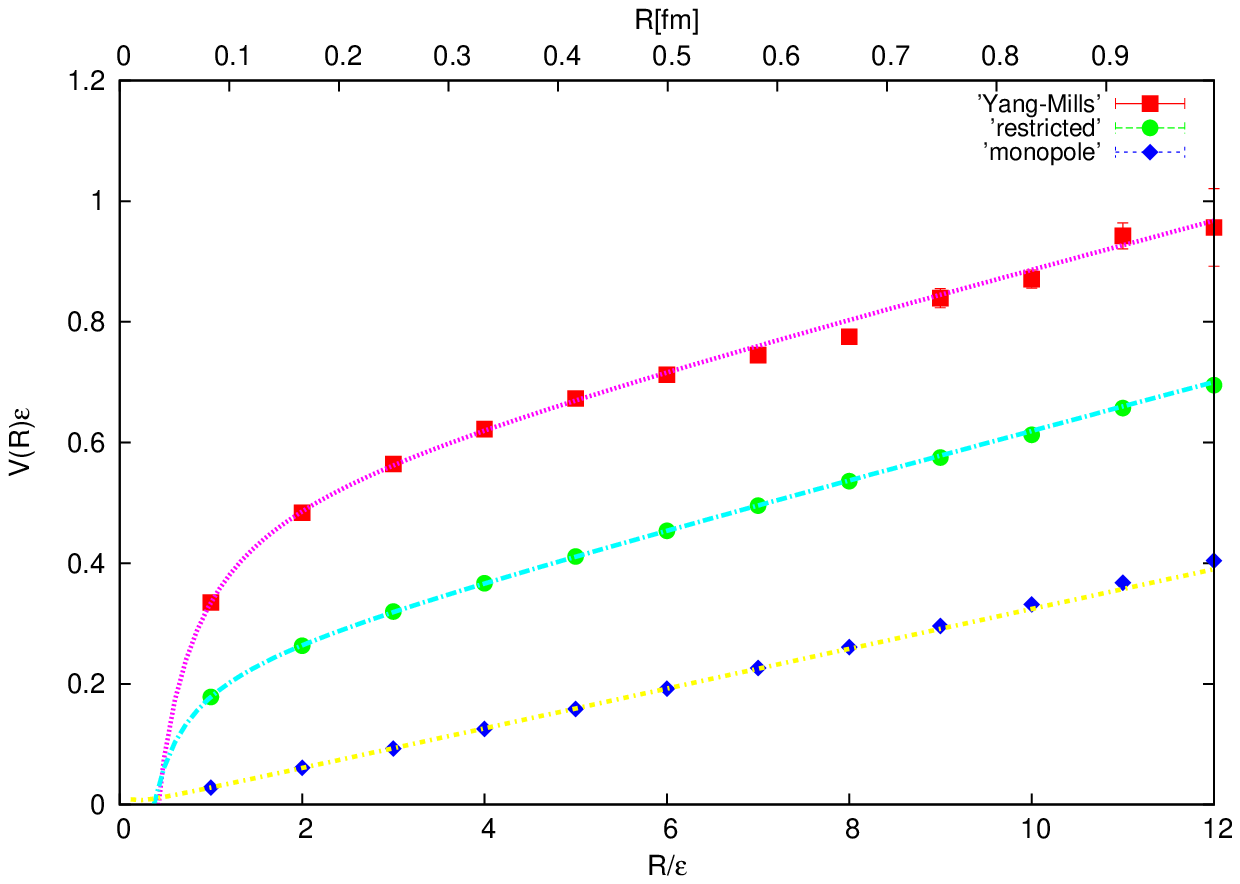}
\quad 
\includegraphics[scale=0.50]{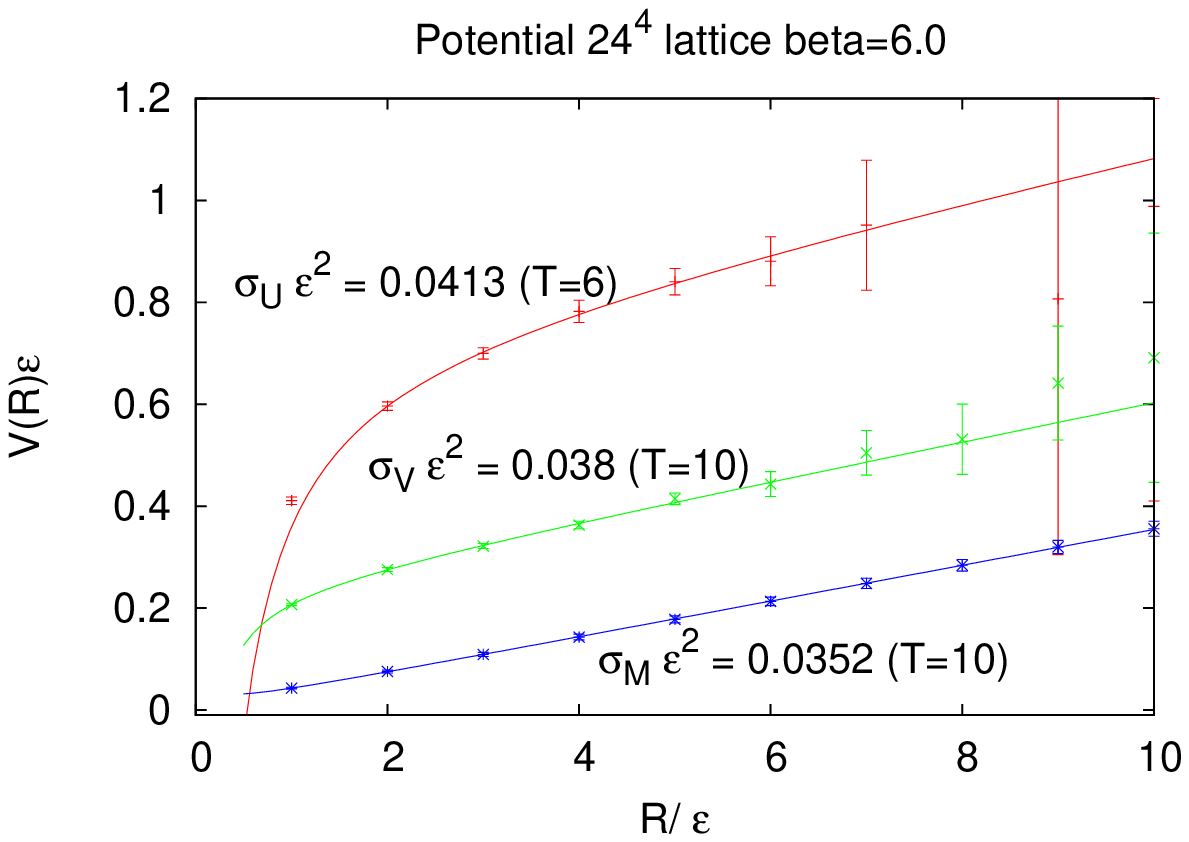}
\end{center}
\vspace{-5mm}
\caption{
The static quark--antiquark potential $V(R)$ as a function of the  distance $R$ in $SU(N)$ Yang-Mills theory.
%, which is obtained by numerical simulations in the framework of lattice gauge theory.
(from above to below):
The full  potential $V_{\rm full}(R)$,  restricted (or ``Abelian'')  part $V_{\rm rest}(R)$ and magnetic--monopole part  $V_{\rm mono}(R)$.
(Left)  $SU(2)$  at $\beta=2.5$ on $24^4$ lattice~$^{8}$, %\cite{KKS15}, 
(Right)  $SU(3)$ at $\beta=6.0$ on $24^4$ lattice~$^{9}$. %\cite{SKKS13}.
%on $16^4$ lattice at $\beta=2.4$,
%(Right)  
%on $24^4$ lattice at $\beta=2.5$ 
%where the Wilson loop with $T=12$ was used for obtaining  $V_{\rm full}(R)$ and $V_{\rm rest}(R)$, and $T=8$ for  $V_{\rm mono}(R)$. 
}%
\label{fig:potential}
\end{figure}

To begin with, let me summarize the current status of numerical simulations of the \textbf{static quark--antiquark potential}. 
Fig.~\ref{fig:potential} shows that the static quark--antiquark potential $V_{q\bar q}(r)$ as a function of the quark--antiquark distance $r$  
%\cite{KKS14} for $SU(2)$ and \cite{SKKS11} for $SU(3)$. 
%\noindent
%The numerical simulations exhibit that 
%The static quark-antiquark potential $V_{q\bar q}(r)$ 
is well fitted by the form of the \textbf{Cornell type}: 
Coulomb+{Linear} (See Fig.~\ref{fig:potential})
%You see that the static quark-antiquark potential has the form:
%$V_{q\bar q}(r)$=Coulomb+{ Linear}:
\begin{equation}
 V_{q\bar q}(r) = - \frac{\alpha}{r} + { \sigma r } + c 
%\rightarrow \infty \ (r \rightarrow \infty)  \Longrightarrow \textbf{quark confinement} 
 ,
\nonumber
\end{equation}
where the parameters have the following dimensions, 
the Coulomb coefficient $\alpha=$  [mass$^0$],  the string tension $\sigma=$  [mass$^2$],  and a constant $c=$   [mass$^1$]. 
Therefore, the potential goes to infinity $V_{q\bar q}(r) \to \infty$ as the distance increases $r \to \infty$, leading to \textbf{quark confinement}. 
%\\
%$\bullet$ $\sigma \not= 0$ confinement $V_{q\bar q}(r) \rightarrow \infty$ as $r \rightarrow \infty$
%\\
%$\bullet$ $\sigma = 0$ deconfinement $V_{q\bar q}(r) < \infty$ as $r \rightarrow \infty$

\begin{figure}[ptb]
%\begin{center}
%%\vspace{-5mm}%
\includegraphics[scale=0.5]{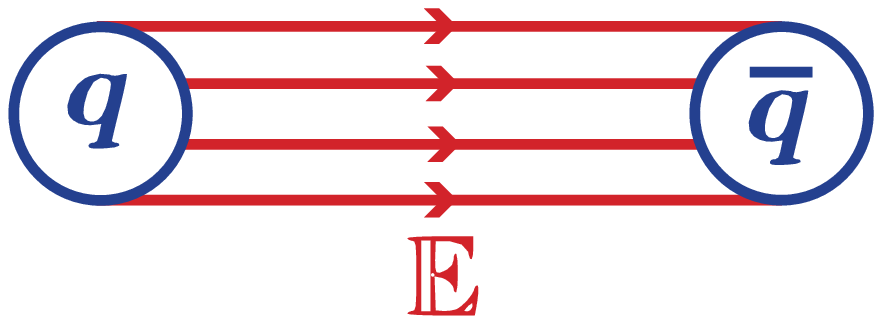}
\quad $\leftarrow$ dual $\rightarrow$ \quad 
\includegraphics[scale=0.5]{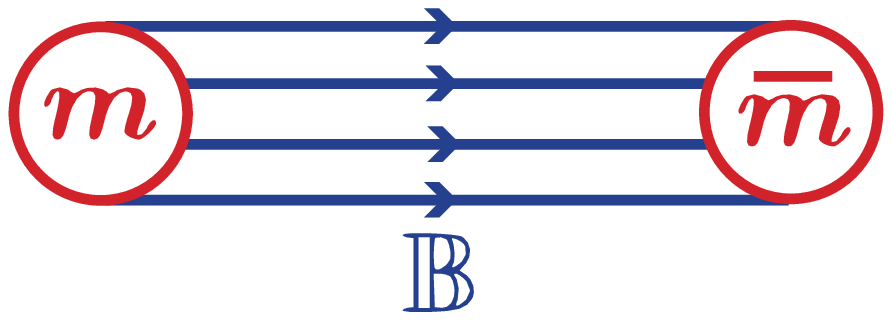}
%%\vspace{-9mm}
%\end{center}
\caption{
The electro-magnetic duality: electric charge is replaced by the magnetic charge, and the electric field is replaced by the magnetic field, and vice versa. 
%\quad\quad\quad\quad\quad\quad\quad
% (Left panel)  
%dual superconductor 
%\quad\quad\quad\quad
%(Right panel) 
%\quad\quad
% superconductor
}
\label{fig:duality}%
\end{figure}

A promising scenario for understanding quark confinement is  the 
\textbf{dual superconductor hypothesis for quark confinement} based on the electro-magnetic duality (See Fig.~\ref{fig:duality}) proposed by 
 Nambu (1974), 't Hooft (1975), Mandelstam (1976)~$^{1}$ %\cite{dualsuper}, 
and Polyakov (1975,1977)~$^{2}$.% \cite{Polyakov75}.
The key ingredients for the dual superconductivity are as follows. 
For reviews, see reviews~$^{3-5}$.%\cite{CP97,Greensite03,KKSS15}. 
%QCD vacuum= %are will  be confirmed by showing  
\begin{itemize}
\item
\textbf{dual Meissner effect}% (See Shibata's talk for the numerical confirmations on a lattice)
%%\vskip -0.5cm
\\
In the dual superconductor, the chromoelectric flux must be squeezed into  tubes.
{ [$\leftarrow$  In the ordinary superconductor (of the type II), the magnetic flux is squeezed into tubes.]} 
%(See Fig.~\ref{fig:glue-string})
%(no chromomagnetic field)

\item
\textbf{condensation of chromomagnetic monopoles  }
\\
The dual superconductivity will be caused by condensation of magnetic monopoles (called the chromomagnetic monopoles).
{
[$\leftarrow$ The ordinary superconductivity is cased by  condensation of electric charges into Cooper pairs.]}

\end{itemize}

%%%%%%%%%%%%%%%%%%%%%%%%%%%%%%%%%%%%%%%%%%%%%%%%%%%%
%%%%%%%%%%%%%%%%%%%%%%%%%%%%%%%%%%%%%%%%%%%%%%%%%%%%

%\newpage
%%%%%%%%%%%%%%%%%%%%% figures %%%%%%%%%%%%%%%%%%%%%%%%%%%
%%%%%%%%%%%%%%%%%%%%% figures %%%%%%%%%%%%%%%%%%%%%%%%%%%

%\noindent
In order to establish the dual superconductivity, we must answer the following questions:
% For this, we must introduce the magnetic monopole ...
%\noindent
%\textbf{\Large Duality and magnetic-monopole in Yang-Mills theory}
%%\vskip -0.5cm

\begin{itemize}
\item
\noindent
How to introduce \textbf{chromomagnetic monopoles} in the Yang-Mills theory without scalar fields?  
[This should be compared with the 't Hooft-Polyakov magnetic monopole.]

\item
\noindent
How to define the \textbf{electric-magnetic non-Abelian duality} in the non-Abelian gauge theory?

%\item
%\noindent
%* How to preserve the {   original (non-Abelian) gauge symmetry}?

\item
\noindent
How to extract the \textbf{infrared dominant mode} %$\mathscr{V}$ 
for confinement?

\end{itemize}

%\newpage
In this talk,  
\begin{itemize}
\item
We give a \textbf{gauge-invariant definition of (chromo)magnetic monopoles}  in the $SU(N)$ Yang-Mills theory (in the absence of the scalar fields) from the non-Abelian Wilson loop operator. 
This is achieved by using a \textbf{non-Abelian Stokes theorem} for the Wilson loop operator~$^{6}$. % \cite{Kondo08}.
This leads to the \textbf{non-Abelian magnetic monopoles} for the Wilson loop operator in the fundamental representation. 
This definition is independent of the gauge fixing. 
One does not need to use the conventional prescription called the \textbf{Abelian projection} proposed by  't Hooft (1981)~$^{7}$ %\cite{tHooft81} 
  which realizes magnetic monopoles by a partial gauge fixing as \textbf{gauge-fixing defects}.

\end{itemize}

%%%%%%%%%%%%%%%%%%%%%%%%%%%%%%%%%%%%%%%%%%%%%%%%%%%%
%\begin{figure}[ptb]
%\begin{center}
%\includegraphics[scale=0.7]{fig-ep206/v_ebi_vs_y_l24c500_ppt.eps} 
%{figs/v_ei_vs_y.eps} 
%\quad
%\includegraphics[scale=0.8]{fig-ep206/v_Ez_3d_hyp2h_c500.eps} 
%\vspace{-5mm}
%\end{center}
%\caption{\small
%The chromoelectric field obtained from the restricted field $V$ on  $24^4$ lattice at $\beta=2.5$. 
%}
%\label{cf-fig3}
%\end{figure}
%%%%%%%%%%%%%%%%%%%%%%%%%%%%%%%%%%%%%%%%%%%%%%%%%%%%

%%%%%%%%%%%%%%%%%%%%%%%%%%%%%%%%%%%%%%%%%%%%%%%%%%%%%
\begin{figure}[ptb]
\begin{center}
%\vspace{-5mm}
\includegraphics[scale=0.30]{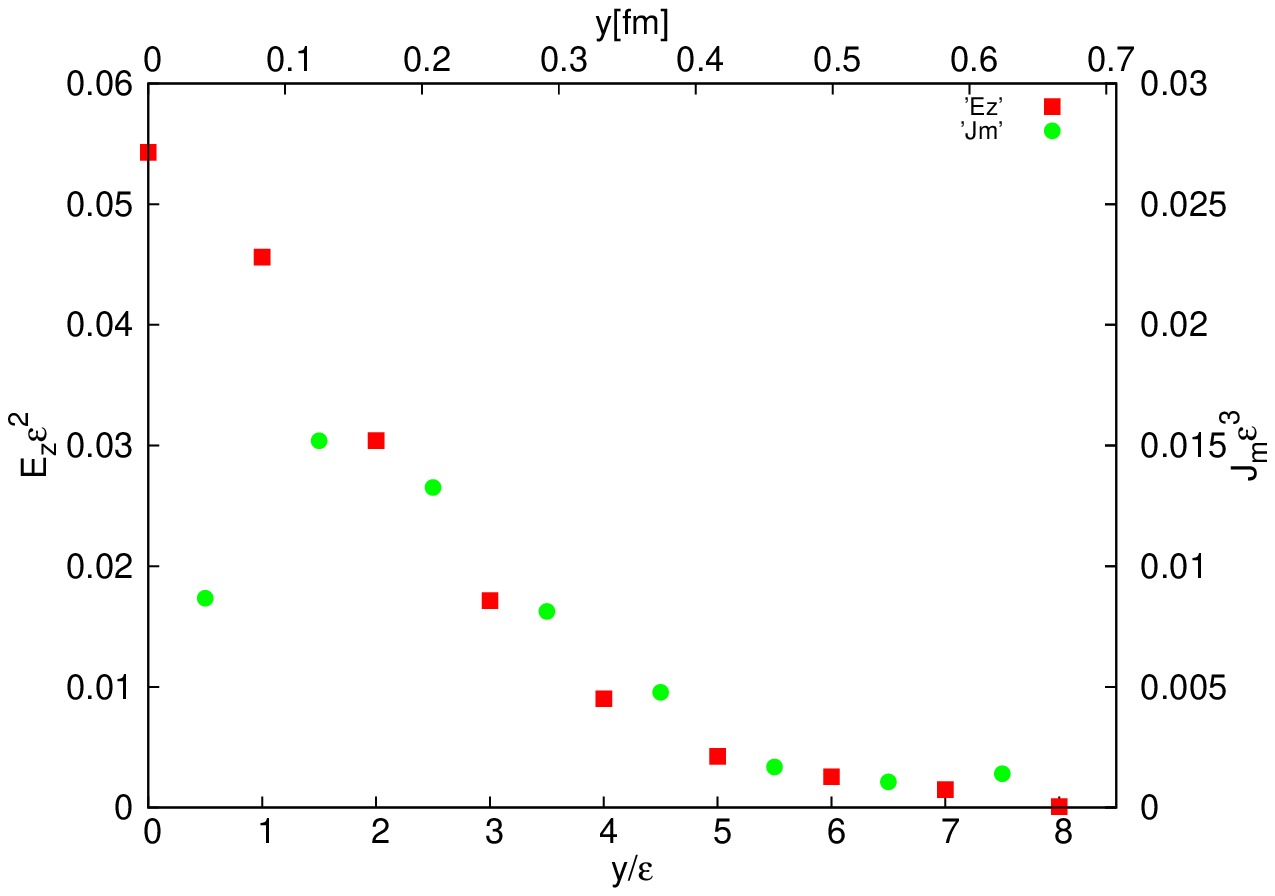}
%{figs/C-flux-M-current.eps} 
\includegraphics[scale=0.65]{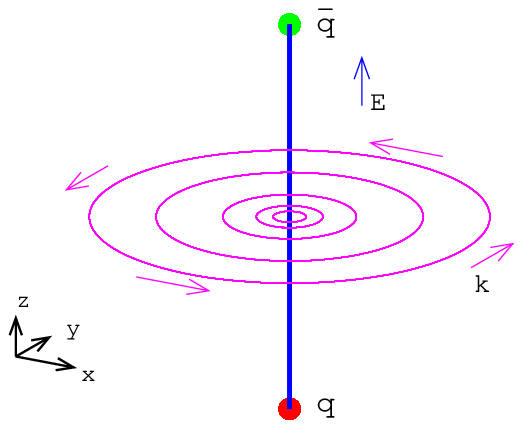} 
\includegraphics[scale=0.30]{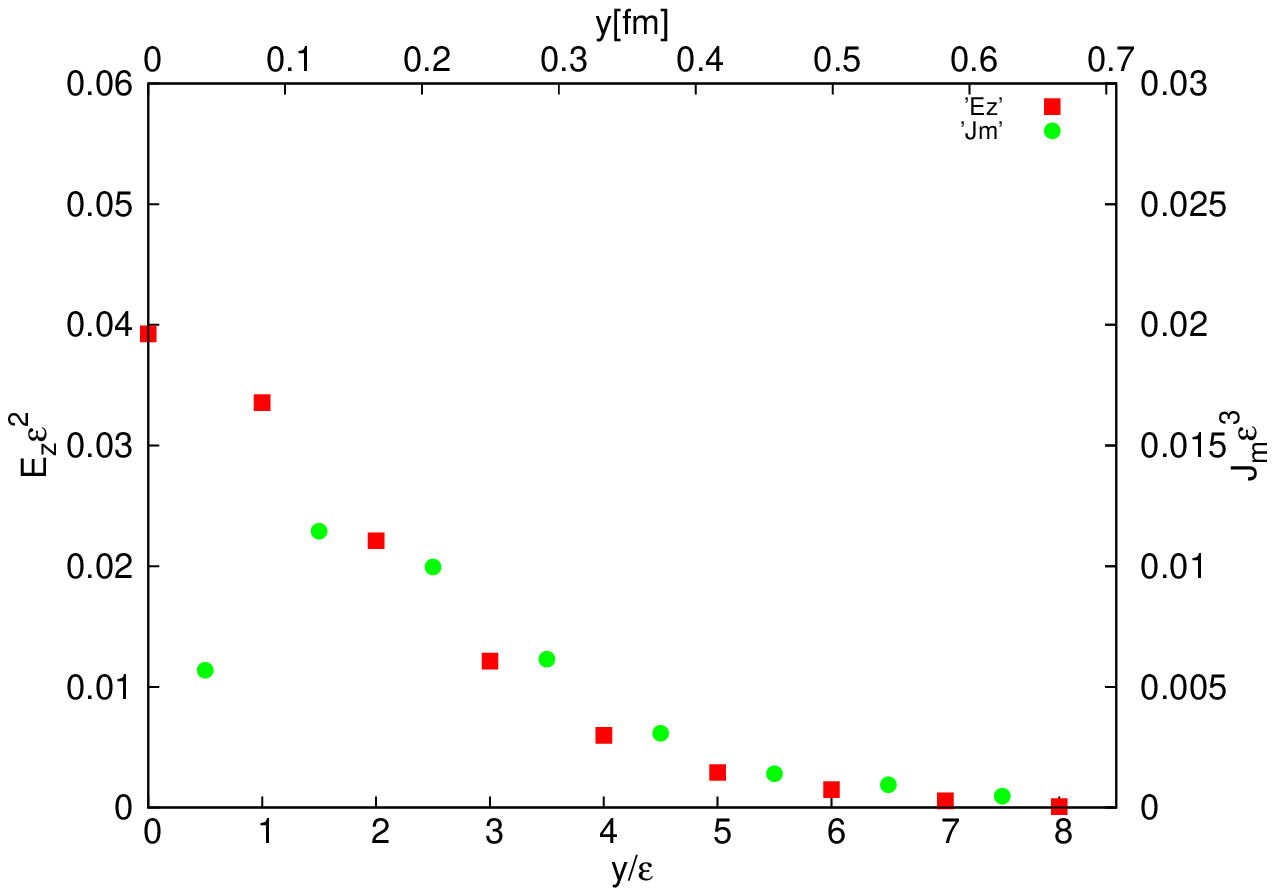}
\vspace{-5mm}
\end{center}
\caption{Ref.$^{10}$: %{} \small
The magnetic-monopole current $\mathbf{k}$ induced around the chromoelectric flux along the $z$ axis connecting a  pair of quark and antiquark.
(Center panel) 
The positional relationship between the chromoelectric field $E_{z}$ and the magnetic current $\mathbf{k}$. 
(Left panel) 
The magnitude of the chromoelectric field $E_{z}$ and the magnetic current  $J_{m}=|\mathbf{k}|$ as functions of the distance $y$ from the $z$ axis calculated from the original full variables. 
(Right panel) 
The counterparts of the left graph calculated from the restricted variables. 
}
\label{fig:Mcurrent}%
\end{figure}
%%%%%%%%%%%%%%%%%%%%%%%%%%%%%%%%%%%%%%%%%%%%%%%%%%%%%

%%%%%%%%%%%%%%%%%%%%% figures %%%%%%%%%%%%%%%%%%%%%%%%%%%
\begin{figure}[ptb]
\begin{center}
\includegraphics[
width=4.0cm,
angle=270
]
{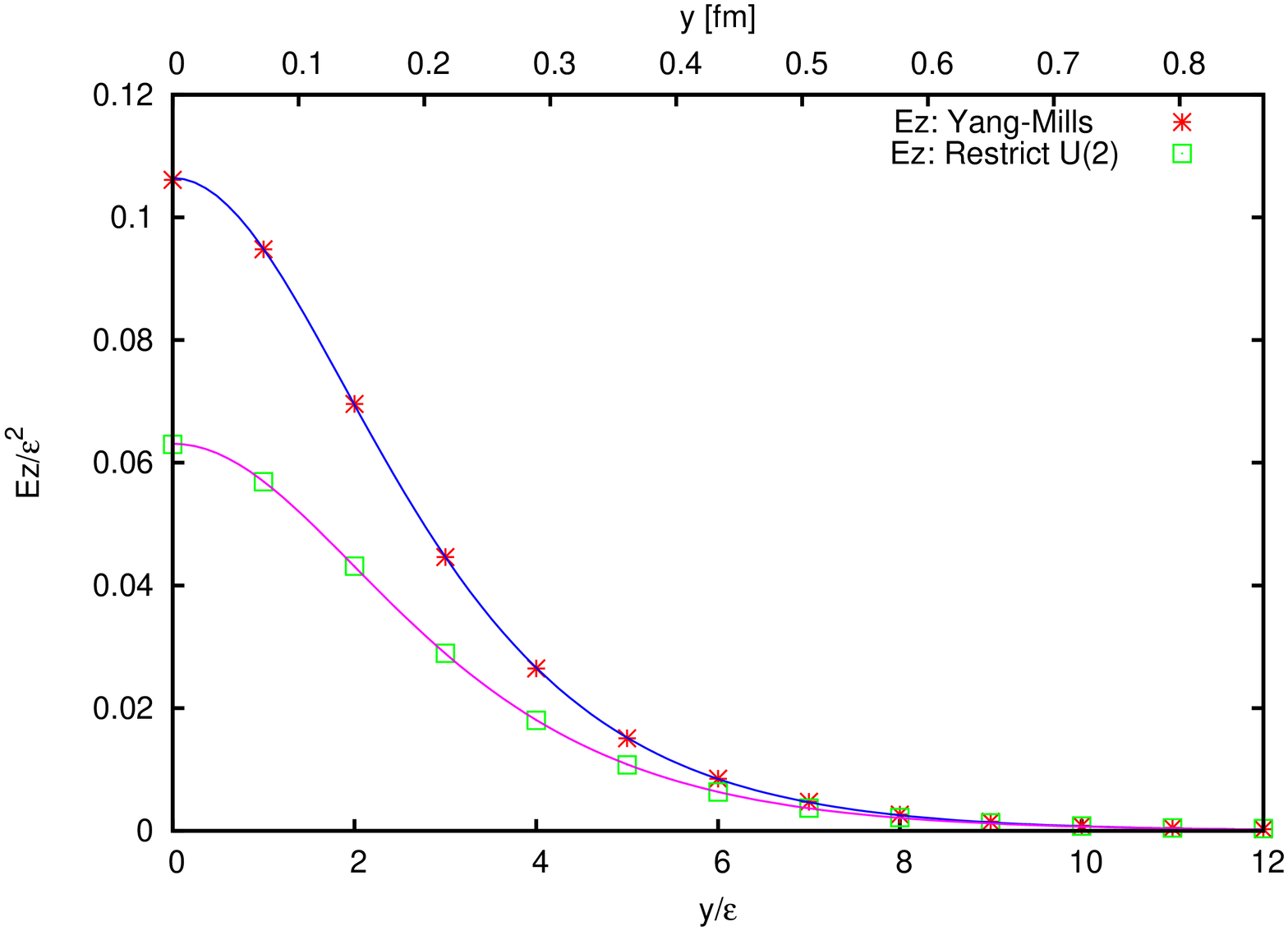} 
\includegraphics[,
width=4.0cm,
angle=270
]
{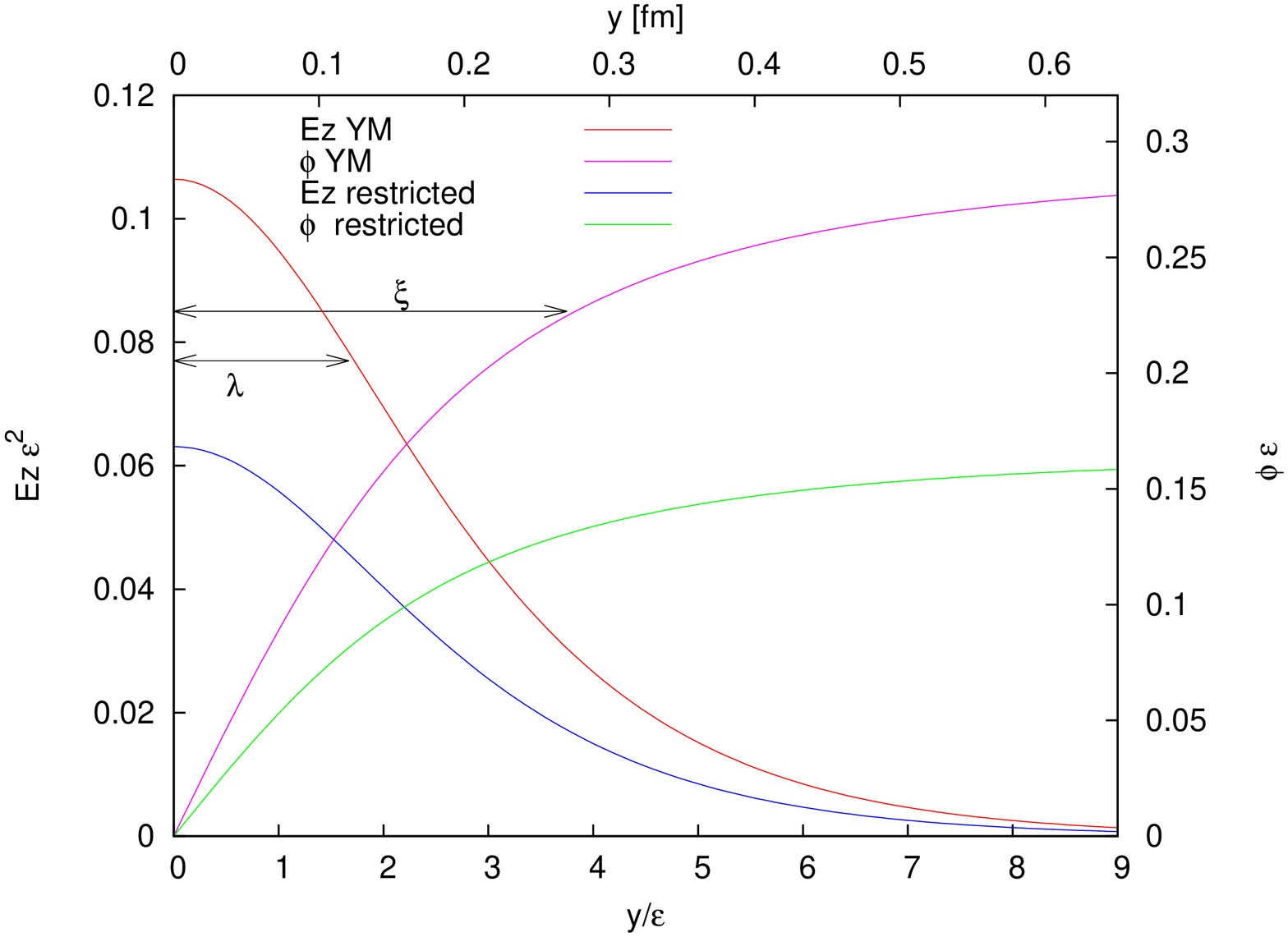}  
\vspace{-0.5cm}
\end{center}
\caption{Ref.~$^{10}$: %\cite{SKKS13} 
%\small
(Left panel)
The plot of the chromoelectric field $E_z$ versus the distance $y$ in units of the lattice spacing $\epsilon$ and the fitting   as a function $E_z(y)$ of  $y$. % according to (\ref{C35-Clem-fit}). 
The red cross for the original $SU(3)$ field and the green square symbol for the restricted field.  
(Right panel) The order parameter $\phi$ reproduced as a function $\phi(y)$ of  $y$. % according to (\ref{order-f}), together with the chromoelectric field $E_z(y)$.
}
\label{C35-fig:type}%
\end{figure}
%%%%%%%%%%%%%%%%%%%%% figures %%%%%%%%%%%%%%%%%%%%%%%%%%%

In fact, the validity of this definition has been confirmed by numerical simulations on a lattice  as follows.
\begin{itemize}
\item
%We have confirmed by numerical simulations on a lattice that  
The magnetic monopole reproduces the linear potential with almost the same string tension $\sigma_{\rm mono}$ as the original one $\sigma_{\rm full}$: 
85\% for $SU(2)$~$^{8}$ %\cite{KKS15}, 
80\% for $SU(3)$~$^{9}$ %\cite{SKKS11}.
This is called the \textbf{magnetic monopole dominance} in the string tension. 

\item
%We have confirmed by numerical simulations on a lattice that  
The \textbf{dual Meissner effect} occur in $SU(N)$ Yang--Mills theory as signaled by the simultaneous formation of the \textbf{chromoelectric flux tube} and the \textbf{associated magnetic-monopole current} induced  around it. 
%See Fig.~\ref{cf-fig2}.
%\newpage
Only the component $E_{z}$ of the chromoelectric field $(E_x,E_y,E_z)=(F_{14},F_{24},F_{34})$ connecting $q$ and $\bar q$ has non-zero value. 
The other components are zero consistently within the numerical errors. 
%In other words, the chromoelectric field is directed to the line connecting  quark and antiquark.
Therefore, the chromomagnetic field $(B_x,B_y,B_z)=(F_{23},F_{31},F_{12})$ connecting $q$ and $\bar q$ does not exist.  
The magnitude of the chromoelectric field $E_{z}$   decreases quickly as the distance $y$ in the direction perpendicular to the line increases.
%Thus the obtained profile of the chromoelectric field represents the structure expected for the flux tube.
Therefore, we have confirmed the formation of the chromoelectric flux in Yang--Mills theory on a lattice~$^{8,10}$. %\cite{KKS15,SKKS13}.

\item
%The above results show the simultaneous formation of the chromoelectric flux tube and the associated magnetic-monopole current induced  around it.    
%Thus, we have confirmed the dual Meissner effect in $SU(N)$ Yang--Mills theory on a lattice. 
We have also shown that the restricted field $V$ reproduces the dual Meissner effect in the $SU(N)$ Yang--Mills theory on a lattice~$^{8,10}$. %\cite{KKS15,SKKS13}.  

\end{itemize}

%\noindent
The superconductor is characterized by 
 the penetration depth $\delta$, 
the coherence length $\xi$, and 
 the Ginzburg-Landau (GL) parameter  $\kappa$
defined  by
\begin{align} 
\kappa := \frac{\delta}{\xi} 
%<\frac{1}{\sqrt{2}}   (\text{type I}), 
%\  =\frac{1}{\sqrt{2}}   (\text{BPS}),
%\  >\frac{1}{\sqrt{2}}  (\text{type II}) 
\begin{cases}
 <\frac{1}{\sqrt{2}} & (\text{type I})\cr
 =\frac{1}{\sqrt{2}} & (\text{BPS}) \cr
 >\frac{1}{\sqrt{2}} & (\text{type II})
\end{cases}
.
 %= \frac{\lambda}{\xi} .
\label{cf4-1}
\end{align}
%The superconductor is called the type-I when $\kappa<\frac{1}{\sqrt{2}}$, while type-II when $\kappa>\frac{1}{\sqrt{2}}$. The border $\kappa=\frac{1}{\sqrt{2}} \simeq 0.707$ is called the BPS limit. 
In the type-I superconductor, the attractive force acts between two flux tubes, while the repulsive force in the type-II superconductor.
There is no interaction at $\kappa=\frac{1}{\sqrt{2}} \simeq 0.707$. 

In the Abelian-Higgs model, two vortices are attractive in the type I,  repulsive for the type II, and there are no force between two vortices in the BPS limit. 
In the Abelian case, therefore, the ordinary \textbf{superconductivity} must be type II or BPS.  Otherwise, the initial configuration of the vortices will be eventually collapsed by the attractive force among the vortices. 
In fact, this is consistent with a fact that the vortices of magnetic flux tube form the \textbf{Abrikosov lattice} (hexagonal lattice rather than forming a square lattice) in the superconductor as stable configuration, which is verified experimentally.

%However, the  numerical simulations show that the dual superconductivity of the Yang-Mills vacuum is \textbf{type I}, in contrast to the preceding studies which claim the border between type I and type II, i.e., BPS limit. $\kappa_{c}=1/\sqrt{2}\simeq0.707$
In the non-Abelian case, recent numerical simulations show that the dual superconductivity of the Yang-Mills vacuum is \textbf{type I}, in contrast to the preceding results which claim the border between type I and type II, i.e., BPS limit. 
%$\kappa_{c}=1/\sqrt{2}\simeq0.707$
\\
\noindent
For $SU(2)$,  Kato et al.(2014)~$^{8}$ %\cite{KKS15} 
reports the GL parameter, the penetration depth, the coherence length:
\begin{align} 
 \kappa = 0.48 \pm 0.07, \ \delta = 0.12{\rm fm}, \ \xi = 0.25{\rm fm} ,
%\nonumber\\
 (m_A=1.64{\rm GeV}, \ m_\phi= 1.1 {\rm GeV})
\end{align}
For $SU(3)$,  Shibata et al.(2013)~$^{10}$. %\cite{SKKS13} 
reports
\begin{align} 
 \kappa = 0.45 \pm 0.01, \ \delta = 0.12{\rm fm}, \ \xi = 0.27{\rm fm} ,
%\nonumber\\
 (m_A=1.64{\rm GeV}, \ m_\phi= 1.0 {\rm GeV})
\end{align}
%The preceding studies support that the dual superconductor for the $SU(2)$ lattice Yang--Mills theory is at the border between type-I and type-II, or  weak type-I \cite{Suzuki:1988}.
These results are obtained using the novel reformulation of the Yang-Mills theory, which is  Cho-Duan-Ge decomposition$^{11-17}$ for $SU(2)$, and Cho-Faddeev-Niemi decomposition$^{18-20}$ for $SU(3)$.

%%%%%%%%%%%%%%%%%%%%%%%%%%%%%%%%%%%%%%%%%%%
%%%%%%%%%%%%%%%%%%%%%%%%%%%%%%%%%%%%%%%%%%%
\section{Non-Abelian vortex}
%%%%%%%%%%%%%%%%%%%%%%%%%%%%%%%%%%%%%%%%%%%
%%%%%%%%%%%%%%%%%%%%%%%%%%%%%%%%%%%%%%%%%%%

The \textbf{type I dual superconductivity} for the Yang-Mills vacuum yields the attractive force between two non-Abelian vortices. 
How this result is consistent with the above considerations?
The crucial point is that the \textbf{non-Abelian vortices} have internal degrees of freedom, i.e., \textbf{orientational moduli}, in addition to the degrees of freedom related to the positions in space. 
This has a possibility for preventing the vortices from collapse due to the attractive force.  
Whereas the Abelian vortices have only the positions and the collapse of the lattice structure for the Abelian vortices in the type I superconductor will not be avoidable. 
We must examine the interaction among vortices depending on the orientational moduli in more detail. 
This analysis gives an  estimate of the string tension based on the vortex condensation picture, and possible interactions between two non-Abelian vortices.

We discuss how the \textbf{magnetic monopole condensation} picture are compatible with the \textbf{vortex condensation} picture as another promising scenario for quark confinement. 
%We analyze the profile function of the magnetic flux tube as the non-Abelian vortex solution of $U(N)$ gauge-Higgs model, which is to be compared with numerical simulations on a lattice for $SU(N)$ Yang-Mills theory.
As a dual or complementary point of view, we can define a gauge-equivalent thin vortex (non-orientable) which has the magnetic monopole and the anti-magnetic monopole as the boundaries~$^{21}$. %\cite{Kondo08b}.

We want to construct the non-Abelian vortex which ends at the  non-Abelian magnetic monopole, that is to say, the non-Abelian orientational zero modes of the vortex endow the endpoint non-Abelian magnetic monopole and antimonopole with same $CP^{N-1}$ zero modes. 

Such a vortex will be obtained in the Higgs phase of a $U(N)$ gauge theory with $SU(N)$ flavor symmetry   where the vortices with non-Abelian $CP^{N-1}$ orientational zero modes exist. 
This is a candidate of the dual gauge theory for describing the magnetic flux tube as a vortex solution. 
This is different from the dual Abelian-Higgs model with the magnetic gauge symmetry $U(1)^{N-1}$ suggested from the Abelian projection.

%%%%%%%%%%%%%%%%%%%%%%%%%%%%%%%%%%%%%%%%%%%
%%%%%%%%%%%%%%%%%%%%%%%%%%%%%%%%%%%%%%%%%%%
\section{Vortex picture towards the area law}
%%%%%%%%%%%%%%%%%%%%%%%%%%%%%%%%%%%%%%%%%%%
%%%%%%%%%%%%%%%%%%%%%%%%%%%%%%%%%%%%%%%%%%%

Then we discuss how such magnetic monopoles can give a nontrivial contribution to the Wilson loop operator for understanding the area law of the Wilson loop average.

The vortex condensation picture gives an easy way to understand the area law. 
Let us assume that the vacuum is filled with percolating thin vortices.
Suppose that $N$ random vortices intersect a plane of area $L^2$.  %See Fig.~\ref{C29-fig:vortex-picture}.
For simplicity, we consider the $SU(2)$ gauge group in what follows.
Each intersection multiplicatively contributes a factor $(-1)^{2J}$ to the Wilson loop average 
and $n$ intersect within the loop take the value $(-1)^{2Jn}$.  
Then the probability that $n$ of the intersections occur within an area $S$ spanned by a Wilson loop is given by
$(-1)^{2Jn} \left( \frac{S}{L^2} \right)^{n} (+1)^{N-n} \left( 1-\frac{S}{L^2} \right)^{N-n}$. 

Summing over all possibilities with the proper binomial weight yields 
\begin{align}
  W_{C}   
&= \sum_{n=0}^{N} 
\begin{pmatrix}
N \\
n 
\end{pmatrix}
(-1)^{2Jn} \left( \frac{S}{L^2} \right)^{n} (+1)^{N-n} \left( 1-\frac{S}{L^2} \right)^{N-n}
 \nonumber\\
&= \left( 1-\frac{S}{L^2} + (-1)^{2J} \frac{S}{L^2}  \right)^{N}
 \nonumber\\
&= \left( 1- \frac{[1-(-1)^{2J}]\rho S}{N}  \right)^{N}
 \to 
 \exp \left\{ -\sigma_J S \right\} \quad (N \to \infty),
 \quad
 \rho := \frac{N}{L^2}  ,
\end{align}
where
\begin{align}
\sigma_J = [1-(-1)^{2J}]\rho
= \begin{cases}
  \sigma_F = 2\rho & (J=\frac12 , \frac32, \cdots) \cr
 0 & (J=1,2,\cdots) 
 \end{cases}
 ,
\end{align}
where $L$ has been eliminated in favor of the planar vortex density $\rho:=N/L^2$. 
The limit of a large $N \to \infty$ is taken with a constant $\rho$.  Thus one obtains an area law for the Wilson loop average with the string tension $\sigma_J$ determined by the vortex density $\rho$. 

The crucial assumption in this argument is the independence of the intersection points. 
%Let us assume that the vacuum is filled with percolating thin vortices.
%Divide the minimal surface bounded by the loop $C$ into small patches with the area $\epsilon^{2}$. 
%Let $p$ denote the probability that a patch is intersected by a vortex. We assume that intersectionz are random and uncorrelated. 
%Then it is easy to show that 
%\begin{align}
%& W_{C}   = [ (-1)^{2J} p + (+1)(1-p) ]^{\text{Area}(C)/\epsilon^2}
%= \exp [- \sigma_J \text{Area}(C)],
%\\
%& \sigma_J = - \epsilon^{-2} \ln [(-1)^{2J} p  + 1-p ] 
%= \begin{cases}
% - \epsilon^{-2} \ln [1-2p ] \cong 2p \epsilon^{-2} & (J=\frac12 , \frac32, \cdots) \cr
% 0 & (J=1,2,\cdots) 
%\end{cases}
% ,
%\end{align}
The asymptotic string tensions are zero for all integer-$J$ representations (with $N$-ality or ``biality''  being equal to $0$), while they are nonzero and equal for all half-integer $J$  representations (with $N$-ality or ``biality''  being equal to $1$). 
%This is the asymptotic string tension. See section \ref{sec:Casimir-scaling} for the string tension in the intermediate distance region.
For the details, see Ref.~$^{22}$. %\cite{KSS15}.  

%%%%%%%%%%%%%%%%%%%%%%%%%%%%%%%%%%%%%%%%%%%
%%%%%%%%%%%%%%%%%%%%%%%%%%%%%%%%%%%%%%%%%%%
\section{Conclusion and discussion}
%%%%%%%%%%%%%%%%%%%%%%%%%%%%%%%%%%%%%%%%%%%
%%%%%%%%%%%%%%%%%%%%%%%%%%%%%%%%%%%%%%%%%%%

\begin{enumerate}
\item 
{ [Magnetic monopole picture] }
We have given a gauge-invariant definition of the magnetic monopole in the $SU(N)$ pure Yang-Mills theory in the absence of the scalar field.  
This is achieved through a non-Abelian Stokes theorem for the Wilson loop operator.
This enables one to estimate the magnetic monopole contribution to the Wilson loop average, 
which confirms the magnetic monopole dominance in the string tension.
See Shibata's talk for the numerical simulations on a lattice. 

%\item
%{ [thin vortex] }
%As a dual or complementary point of view, we can define a gauge-equivalent thin vortex (non-orientable) which has the magnetic monopole and the anti-magnetic monopole as the boundaries.

\item
{ [Vortex picture] }
We have discussed the vortex solution in the Higgs phase of a gauge-Higgs model with  $U(N)$ gauge fields and $N$ Higgs fields. 
The model fulfills the requirement that a non-Abelian vortex has the same $\bm{C}P^{N-1}$ moduli space as those of the non-Abelian magnetic monopole. 
The non-Abelian magnetic monopole is regarded as a kink making a junction with $N$ non-Abelian vortices with different internal orientation moduli. 
The internal moduli will be important to understand the type I dual superconductivity.
%\\
%Such vortex solutions will give complementary understanding on confinement. 
%the magnetic flux tube has the magnetic monopole as the endpoints. 

%\item
%Such vortex solutions are expected to give more systematic studies on confinement. 

\end{enumerate}

%%%%%%%%%%%%%%%%%%%%%%%%%%%%%%%%%%%%%%%%%%%
%%%%%%%%%%%%%%%%%%%%%%%%%%%%%%%%%%%%%%%%%%%
\subsection*{Acknowledgement}
%%%%%%%%%%%%%%%%%%%%%%%%%%%%%%%%%%%%%%%%%%%
%%%%%%%%%%%%%%%%%%%%%%%%%%%%%%%%%%%%%%%%%%%

This work is supported by Grant-in-Aid for Scientific Research (C) 24540252
from the Japan Society for the Promotion Science (JSPS), and also in part by JSPS
Grant-in-Aid for Scientific Research (S) 22224003. 
This work is in part supported by the Large Scale Simulation Program 
No.09-15 (FY2009), No.T11-15 (FY2011), No.12/13-20 (FY2012-2013) and No.13/14-23 (FY2013-2014) of High Energy Accelerator Research Organization (KEK).

\end{document}